\documentstyle[12pt]{article}
\newcommand{\newsection}[1]{\vspace*{5mm} \noindent {\bf #1 }
 
\vspace*{-3.0mm}
\noindent}

\newcommand{\DS}{\renewcommand{\baselinestretch}{2} \tiny \normalsize }
\newenvironment{references}{\vspace*{5mm} \noindent {\bf References} \\
\vspace*{-5mm} \small
\begin{description}}{\end{description}}
  
\begin{document}
\DS

\begin{center}
\Large {\bf The Interpretation of Near-Infrared Star Counts at the South 
Galactic Pole}
\vspace*{3.0mm}
 
\large
\large
Takeo Minezaki\\
\normalsize
National Astronomical Observatory, Mitaka-shi, Tokyo 181, Japan and Department of Astronomy, School of Science, The University of Tokyo;
current address: Kiso Observatory, Institute of Astronomy, Faculty of Science, The University of Tokyo,
Mitake-mura, Kiso-gun, Nagano-ken, 397-01, Japan\\
\small
Electronic mail:  minezaki@kiso.ioa.s.u-tokyo.ac.jp

\large
Martin Cohen\\
\normalsize
Radio Astronomy Laboratory, 601 Campbell Hall, University of California, Berkeley, CA 94720
\& Radio Astronomy Laboratory, 601 Campbell Hall, University of California,
Berkeley, CA 94720, and Vanguard Research, Inc., Suite 204, 5321 Scotts Valley Drive, 
Scotts Valley, CA 95066 \\
\small
Electronic mail:  mcohen@astro.berkeley.edu

\large
Yukiyasu Kobayashi\\
\normalsize
National Astronomical Observatory, Mitaka-shi, Tokyo 181, Japan\\
\small
Electronic mail:  yuki@merope.mtk.nao.ac.jp

\large
Yuzuru Yoshii\\
\normalsize
Institute of Astronomy, Faculty of Science, The University of Tokyo, Mitaka-shi, Tokyo 181, Japan
\& Research Center for the Early Universe, School of Science, The
University of Tokyo, Bunkyo-ku, Tokyo 113, Japan\\
\small
Electronic mail:  yoshii@omega.mtk.ioa.s.u-tokyo.ac.jp

\large
Bruce A. Peterson\\
\normalsize
Mt. Stromlo and Siding Spring Observatories, Institute of Advanced Studies, The Australian National University, Private Bag, Weston Creek, A.C.T. 2611, Australia\\
\small
Electronic mail:  peterson@mso.anu.edu.au
\end{center}
 
\begin{abstract}
\noindent
We present new deep $K^\prime$ counts of stars at the South Galactic Pole (SGP) taken 
with the NAOJ PICNIC camera to $K^\prime$=17.25.  Star--galaxy separation 
to  $K^\prime$=17.5 was accomplished 
effectively using image profiles because the pixel size we used is 
0.509 arcsec.  We interpret these counts using the SKY (Cohen 1994) model 
of the Galactic point source sky and determine the relative normalization of 
halo-to-disk populations, and the location of the Sun relative to the 
Galactic plane, within the context of this model.  The observed star
counts constrain these parameters to be: halo/disk 
$\sim$1/900 and z$_\odot$=16.5$\pm$2.5 pc.  These values have
been used to correct our SGP galaxy counts for contamination by the point
source Galactic foreground.
\end{abstract}
\newpage

\newsection{1. Introduction}
 
\noindent
The advent of new near-infrared (NIR) cameras is providing a
wealth of valuable star counts that can be used to constrain
models of the Galaxy.  These models are critical to the quest
for any near- and mid-infrared cosmic background radiation.  
Recently, Minezaki {\it et al.} (1997: hereafter designated as Paper I) 
have presented differential $K^\prime$-band galaxy counts 
very close to the South Galactic Pole (SGP) taken with the PICNIC 
camera using a pixel size of 0.509 arcsec on the 
the Australian National University's 2.3 m telescope
at Siding Spring Observatories, Australia.  The purpose of the
present paper is twofold: to describe our efforts to interpret 
and model the stellar content of these counts through use of
the SKY model (Cohen 1994; Wainscoat {\it et al.} 1992, hereafter
referred to as WCVWS) and 
to justify the manner in which we removed the stellar foreground 
contribution to these galaxy counts in Paper I,
to isolate the extragalactic component. 

We very briefly summarize the relevant observational details of 
Paper I's ``bright survey" in Section 2, and the version of the SKY model 
that we use in Section 3.  In Section 4 we discuss the way in which
we have separated the interpretation of the star counts into
the measurement of the halo/disk normalization factor and of
the location of the Sun with respect to the Galactic plane.
In this we follow precisely the approach taken by Cohen (1995),
who pursued the same two parameters in the context of SKY,
primarily through the use of IRAS source counts at the two Galactic
Poles and of high latitude Shuttle-based far-ultraviolet star counts.  
Our values for these parameters are in good agreement both with 
Cohen's values and those found by conventional techniques and
modeling in the visible.  Section 5 presents an analysis of our
estimated uncertainties in the extrapolation of star counts to faint
$K^\prime$ magnitudes using SKY.\\

\noindent 
\newsection{2. The observations}
\noindent 
A full description of the 
observations, and other analyses, were presented in Paper I.
 The ``bright survey" was carried out during August and September 1994,
 using the Australian National University's 2.3 m telescope
 at Siding Spring Observatory, Australia, equipped with the PICNIC 
near-infrared camera (Kobayashi {\it et al.} 1994),
 developed at National Astronomical Observatory, Japan.
 PICNIC uses a NICMOS3 array ($256\times 256$ pixels)
 with a field of view of $2.2\times 2.2$ arcmin${}^{2}$ 
 and a pixel scale of 0.509 arcsec pixel${}^{-1}$.
 In order to reduce the thermal sky background,
 we used a $K^\prime$ filter, which has the same transmission curve
 as the 2MASS $K_{\rm S}$ filter (McLeod {\it et al.} 1995).

       The survey was centered at B2000 coordinates $(\alpha,\delta)=
 (0^{\rm h}50^{\rm m}48^{\rm s},-27^\circ 43^\prime 34"$)
 or $(l,b)=(316.27^\circ,-89.39^\circ)$, and covered 180.8 arcmin${}^{2}$.
 The FWHM of PSF was 1.5 arcsec and the limiting magnitude with 80\%
 completeness was determined as $K^\prime=19.1$ 
from the simulations of the recovery of stellar profiles 
({\it cf.} Paper I).

      Stars and galaxies in the survey were separated
 to $1.5$ mag brighter than the limiting magnitude, $K^\prime=17.5$,
 based on two morphological parameters, the FWHM and the {\it ir1},
 where the FWHM was measured by Gaussian fitting
 of the radial profile by IRAF {\it imexamine} task
 and the {\it ir1} was the intensity-weighted first moment radius
 which was measured by FOCAS.

     The $K^\prime$ band star counts to $K^\prime=17.25$ obtained 
from the survey are tabulated in Table 1.

\newsection{3. The version of SKY used}

\noindent
We have utilized version 4 of the SKY model, exactly as 
documented by Cohen (1994) and Cohen {\it et al.} (1994).  The
difference between implementations of the ``standard" $K$ and the
newer $K^\prime$ filters in SKY have been explored by Cohen (1997), 
who found differences of at most 0.03 mag 
between the absolute magnitudes, M$_K$ and M$_{K^\prime}$, for
the entire set of 87 categories of source with which SKY
populates its representation of the sky.  These differences are
negligible compared with both the intrinsic dispersions and 
uncertainties in the absolute magnitudes of the many types of
celestial source, and with the photometric uncertainties in the
measurement of $K^\prime$ at faint magnitude levels.  Therefore,
our simulations with SKY were based on SKY's ``hardwired'' standard
$K$ passband.

SKY incorporates five geometric components: the (thin) disk; bulge;
spiral arms, local spurs, and Gould's Belt; molecular ring; and halo.
SKY does not formally include a thick disk ({\it cf.} Yoshii 1992;
Yoshii {\it et al.} 1988), principally because it has negligible
influence on currently available infrared star counts all of whose
Poisson errors greatly exceed the thick disk's expected contribution
in any direction.  

\newsection{4. The SKY simulations}

\noindent
We first motivate the analysis with SKY by a simple analytical approach.
Assuming that the vertical disk scale-height $h_z$ is the same above and 
below the disk plane, the solar displacement z$_\odot$ can be roughly 
evaluated from the ratio between the observed star counts in the north 
and south Galactic poles (hereafter referred to as NGP and SGP): 
$$
\frac{NGP}{SGP}=\frac{e^{-x_\odot}(x_\odot^2+2x_\odot+2)}
                {4-e^{-x_\odot}(x_\odot^2+2x_\odot+2)} \;\;\; ,
$$                
where $x_\odot=z_\odot/h_z$.  We currently lack comparably deep $K^\prime$ 
counts at the NGP to combine with our own SGP infrared star counts.  The
only counts we have examined near the NGP in $K^\prime$ are from the 2MASS
pilot study of SA57 at {\it b}=85$^\circ$ (Skrutskie 1996), to a limit of
14th magnitude.  Using our own and these 2MASS data we can determine only 
that z$_\odot$ is greater than zero, because so few stars are available
to the common limit of 14.

Therefore, we attempted to
derive $x_\odot$ and, hence, $z_\odot$ (since $h_z$ is known for each 
category of source) using the relation
$$
SGP=\omega\rho_0h_z^3\int_{x_\odot}^{\infty}e^{-x}x^2dx
=\omega\rho_0h_z^3(4-e^{-x_\odot}(x_\odot^2+2x_\odot+2)) \;\;\; ,
$$
for one-sided star counts, to constrain $z_\odot$ since 
$\rho_0$ is known.  $\omega$ represents the area in steradians
over which we assess the star counts, and we summed over all categories
of star.

Using values for scale height and space density purely for the thin disk 
from WCVWS (their Table 2), and our observed cumulative counts
(corrected for incompleteness) of 67.17 stars within the bright survey 
area, we were again unable to obtain more than an approximate
value for $z_\odot$ because of the appreciable Poisson 
uncertainties in our counts.

Consequently, we chose to interpret the counts strictly within the
context of SKY in which many stellar populations are represented, along
with the pertinent geometry of halo and disk components.
We compared the counts predicted by SKY with the star counts in Table
1, after applying the corrections for
incompleteness (hereafter referred to as ``corrected counts"), and
after removal of galaxies on the basis of their much
broader photometric profiles.
These counts were binned into 0.5 mag wide bins, then converted to 
differential counts per square degree per magnitude interval.
The sensitivity of our analysis in determining the optimal value of z$_\odot$ 
is less than that associated with halo/disk so we first sought the best 
solution for halo/disk.  We fixed the Sun's position at 15 pc north of the 
Galactic plane ({\it cf.} Cohen 1995) and ran a series of predictions for 
differential counts specific to the $K$-band at
the central coordinates of the mosaic of PICNIC fields, 
varying halo/disk from 0 to 2, in steps of 0.1, where the unit is
the standard ratio of 1/500 adopted by WCVWS.  
Following Cohen (1995), we analyzed three statistics to 
select the best solution for halo/disk and z$_\odot$, namely: unweighted 
algebraic and inverse-variance-weighted average deviations (AD and WAD,
respectively) and the unweighted sum-squared-deviations (SSD).  The former
require the derivation of zero crossings; the latter of an 
absolute minimum.  The three distinct quantities were interpolated 
onto a much finer grid of halo/disk ratio.
The formal $\chi^2$ analysis was again ({\it cf.} Cohen 1995) deemed to be
unsuitable here because the dynamic range of the observations is close to
a factor of 100 in both axes, leading to inappropriately high weighting of
the faintest points observed. Our best fitting values were: WAD, 0.56; 
AD, 0.59; SSD, 0.54 (in units of the standard ratio of 1/500
used by WCVWS).

These procedures lack a natural associated error.
To extract the formal uncertainties we 
assessed the sensitivity of this technique by increasing the 
corrected counts by 
1 star in the faintest 3 (0.5 mag wide) bins, and similarly decreasing 
those in the brightest 3 bins by 1 star.  We then repeated the 
procedure, switching the senses of the increase and decrease.
(Randomly adding or subtracting stars would lead only
to a best-fitting set of counts with more or less the same shape
as that observed, hence the same value of the parameter would result.)
In this way, we derived plausible estimates of the non-systematic 
errors associated with the three statistics as:
WAD, $\pm$0.015; AD, $\pm$0.018; SSD, $\pm$0.031 (strictly, these errors
are slightly asymmetric in the positive and negative directions).  We, therefore,
derive the optimal value for halo/disk as 0.56$\pm$0.03 
in units of the standard ratio of 1/500.

We fixed the ratio halo/disk at this 
formal best-fitting value and similarly constructed the three
statistics for z$_\odot$ by running a series of SKY predictions differing 
by 1 pc in their input value of z$_\odot$, covering the range 5--30
pc.  The identical methodology, 
applied to the analysis of z$_\odot$, yielded: WAD, 14.6; AD, 18.9; 
SSD, 16.0 pc, with uncertainties (assessed in the manner described 
above) of $\pm$3.6, 2.0, and 1.7 pc, respectively.
Therefore, we take 16.5$\pm$2.5 pc as the best estimate for the solar
location using the PICNIC counts and SKY version 4.  

Figure 1 compares this best-fitting simulation with our corrected
star counts.  This simulation was extrapolated to assess the contribution of
faint stars to the total counts in the PICNIC ``bright survey" at 
$K^\prime>$17.5,
below which we did not attempt to separate stars and galaxies.

\newsection{5. Uncertainties in extrapolating counts with SKY}

\noindent
We have attempted to quantify the likely uncertainties in SKY's star
counts as follows.  There are three primary contributors to these
uncertainties; two are traceable directly to the separate influences
of the imprecision with which we know z$_\odot$ and halo/disk.  The
third relates to our knowledge of the space density of the dominant
contributors to the total differential counts at the faintest
$K^\prime$ magnitudes.

To assess the first two elements of uncertainty, we ran SKY for the
four extreme cases in which one of z$_\odot$ and halo/disk was held
at its optimal value (Section 4), while the other was set to its
formal mean$\pm3\sigma$ level.  Consequently, these four
predictions were for (z$_\odot$,halo/disk) of (16.5,0.47), (16.5,0.65),
(9,0.56), and (24,0.56).  From the half-ranges of the difference in
total counts we derived the separate fractional uncertainty
contributions from the two parameters.  Figure 2 includes these
contributions over the relevant $K^\prime$ range.

SKY returns a highly detailed ``log" file of its predictions that
breaks down the total differential counts into the five geometric
components, and each of those into the distinct counts from all the
87 categories of source over any desired range of magnitudes.  From
these files, and on the basis of simulations run against other sets of
observed counts in high latitude fields kindly supplied by Hammersley 
(1996: the North Galactic Pole complete to $H\sim$15) and by 
Skrutskie (1996: for {\it b}=85$^\circ$ from 2MASS in $JHK^\prime$, complete 
to $\sim$16, 15, and 14,
respectively), as well as those obtained by Meadows (1994: complete
to $K^\prime\sim$17.5), (i) we have determined that the ``M LATE V" 
stars fulfil
the role of this dominant population, increasingly so as one 
goes to fainter $K^\prime$ magnitudes;
and (ii) we estimate the likely uncertainty in space
density of these stars at $\pm$0.2 dex.  We can, therefore, derive
the impact upon total counts of a $\pm$0.2 dex change in the space
densities of these cool dwarf stars alone.  This provides the third
component of uncertainty in our analysis (see Figure 2).

All three components are distinct and independent so we combined them,
as fractional uncertainties in a root-sum-square fashion, to yield
our final estimate (the curve labeled ``Total" in Figure 2) of the
$\pm3\sigma$ fractional uncertainty associated with the extrapolation of 
the PICNIC corrected counts to faint magnitudes using SKY's predictions.

By extrapolating the fitted SKY model to $K^\prime>$17.5,
the contributions of the star counts to the total counts were
estimated as about 7\% at $K^\prime$=18, and 5\% at $K^\prime$=19,
which were just comparable to the Poisson errors of galaxy counts
in the PICNIC ``bright survey". Since the estimated uncertainty of 
SKY predictions is about 0.16 dex, or 40\%, at $K^\prime$=18-19,
the uncertainties of the star count predictions would not affect
the galaxy counts of the PICNIC ``bright survey" presented in Paper I.\\

\noindent
\newsection{6. Discussion}
\noindent
There is substantial corroborative evidence in favor of an
offset of the Sun from the plane of order 20 pc from quite
different disciplines (see the references in Cohen 1995), such as the 
distributions of optically-known 
Wolf-Rayet stars (Conti \& Vacca 1990: 15 pc) and of diffuse Galactic
infrared radiation (Hammersley {\it et al.} 1994: 15.5 pc; Arendt {\it et al.} 1996: 
18 pc), all of which are somewhat smaller than the value found from 
analysis of faint red stars (Yamagata \& Yoshii 1992: 40 pc). 
We conclude that our best value, from these 
deep near-infrared counts within the context of the SKY model, is
about 16.5 pc, with a formal 1$\sigma$ uncertainty of about 2.5 pc.

WCVWS originally adopted a halo/disk normalization factor of 1/500 
from Bahcall \& Soneira (1984), within the estimated uncertainties 
obtained by Schmidt (1975), who derived a best value of 1/800 with 
rough probable range 1/550 to 1/1200.  Bahcall {\it et al.}
(1983) similarly deduced ratios between 1/200 and 1/1500, and Cohen 
(1995) suggested a number density ratio of halo/disk of 1/1250 for 
SKY version 4, at the lower bound of Schmidt's suggested approximate
range, but within that determined by Bahcall {\it et al.} (1983). 
The uncertainty inherent in Cohen's (1995) estimate of halo/disk
is probably at least $\pm$0.15, in units of WCVWS's value (1/500).
Within the context of SKY, the PICNIC counts vindicate a value
of halo/disk normalization of 1/900, with probable root-mean-square
range between 1/850 and 1/950, entirely consistent with more traditional 
determinations in the visible by Schmidt, and by Bahcall and colleagues.

There are, of course, issues of uncertainty that strictly fall 
outside the context
of SKY, such as a concern about the completeness of the
complement of sources that SKY incorporates.  Could there, for
example, be a major population omitted from SKY that might
dominate the faint NIR counts, such as red (more extreme than the
SKY category of ``M LATE V"), or brown dwarfs?  While
we cannot preclude this possibility, there are no indications 
of systematic under-predictions by SKY of observed counts at the
faintest levels for which meaningful areas have so far been 
probed, namely $K^\prime\leq$18 (of course, one must always be alert for 
the difficulties inherent in star--galaxy separation and for 
Malmquist bias in faint counts).  Nor is there any evidence for 
the existence of any such cool, faint, halo population from
studies with HST (Gould, Bahcall \& Flynn 1996, 1997). 
Indeed, it might be more fruitful to invert the problem 
and to see which brown dwarf scenarios can already be precluded
by using a model like SKY as an interpreter of existing deep
star counts.\\

\noindent
We thank the SSO staff for their technical support for these
observations, and also thank K. Nakamura for her technical support of 
the observational software.  YY acknowledges financial support from
the Yamada Science Foundation for the PICNIC camera and for the 
Japanese team's observing in Australia.  TM was supported by 
the Grant in Aid for JSPS fellows by the Ministry of Education, Science, 
and Culture. This work has been supported in part by the Grant-in-Aid 
for Center-of-Excellence (COE) Research (07CE2002) of the Ministry of 
Education, Science, and Culture.  The development of SKY has been 
supported largely by Dr. S. D. Price through the sponsorship of 
Phillips Laboratory under contract F19628-92-C-0090, 
with Vanguard Research, Inc. MC also thanks NASA-Ames for 
support under cooperative agreement NCC 2-142 with UC Berkeley.

\newpage

\begin{center}
{\bf TABLE 1.}  The $K^\prime$ band star counts\\

\begin{tabular}{crccc}

$K\prime$ & Raw N$^a$ &  Completeness$^b$ & n$^c$  & Error$^c$\\
\hline
 12.0--12.5 &  1 & 0.998 &  39.9    &  39.9\\
 12.5--13.0 &  1 & 0.998 &  39.9    &  39.9\\
 13.0--13.5 &  7 & 0.998 & 279.  & 106.\\
 13.5--14.0 &  6 & 0.998 & 240.  &  97.8\\
 14.0--14.5 &  3 & 0.998 & 120.  &  69.1\\
 14.5--15.0 &  5 & 0.998 & 200.  &  89.3\\
 15.0--15.5 & 11 & 0.998 & 439.  & 132.\\
 15.5--16.0 &  6 & 0.998 & 240.  &  97.8\\
 16.0--16.5 &  6 & 0.998 & 240.  &  97.8\\
 16.5--17.0 & 10 & 0.998 & 399.  & 126.\\
 17.0--17.5 & 11 & 0.995 & 440.  & 133.\\
\hline
\hline
\end{tabular}
\end{center}

Notes to Table 1:\\
$^a$ Raw counts of detected stars in the specified magnitude range.\\
$^b$ The average of completeness for $15\le K^\prime\le 17$ was presented 
at $K^\prime\le 17$.\\
$^c$ Corrected star counts and the errors per magnitude per degree${}^{2}$.\\

\newpage

\newsection{Figure captions}
\noindent
\vspace*{20mm}

{\bf Figure 1:} Our best match using SKY to the observed PICNIC star counts
for halo/disk of 0.56 $\times$ the 1/500 value used by WCVWS, and 
z$_\odot$ of 16.5 pc. Error bars are the actual Poisson errors
associated with the observed star counts. The curves represent: the
total (differential) counts (solid), and the separate disk (dotted, lighter curve below
the total), and halo (long-dashed--dotted) counts.\\

{\bf Figure 2:} Quantitative estimates of the elements of fractional 
uncertainty associated with the extrapolation of our PICNIC corrected
counts to faint $K^\prime$ levels.  The ``Total" curve is constructed
from the root-sum-square of the 3 separate curves that arise from
uncertainties in our ratio of halo/disk, in the space density of 
late-M dwarf stars, and in the location of the Sun with respect to the
Galactic plane.\\
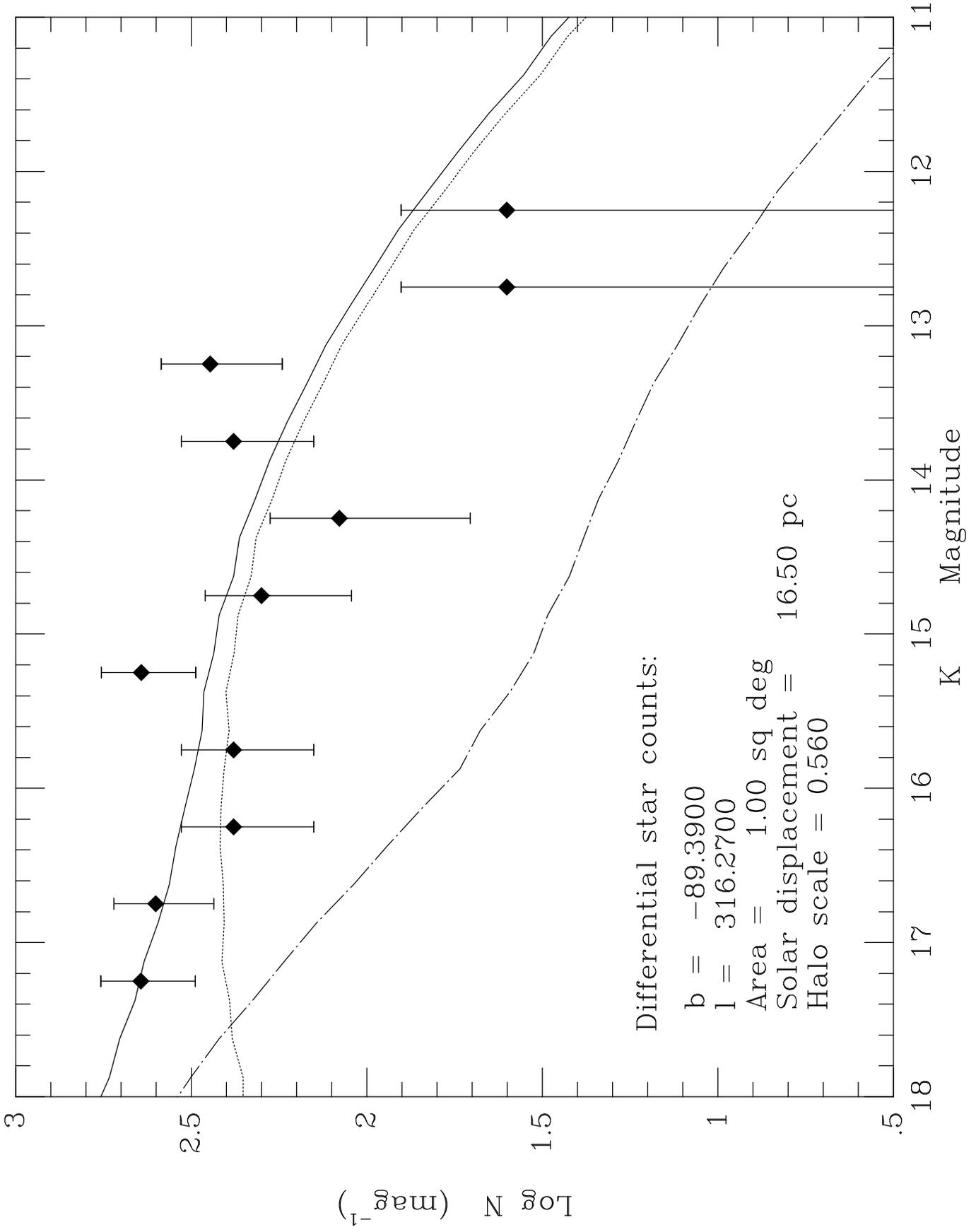
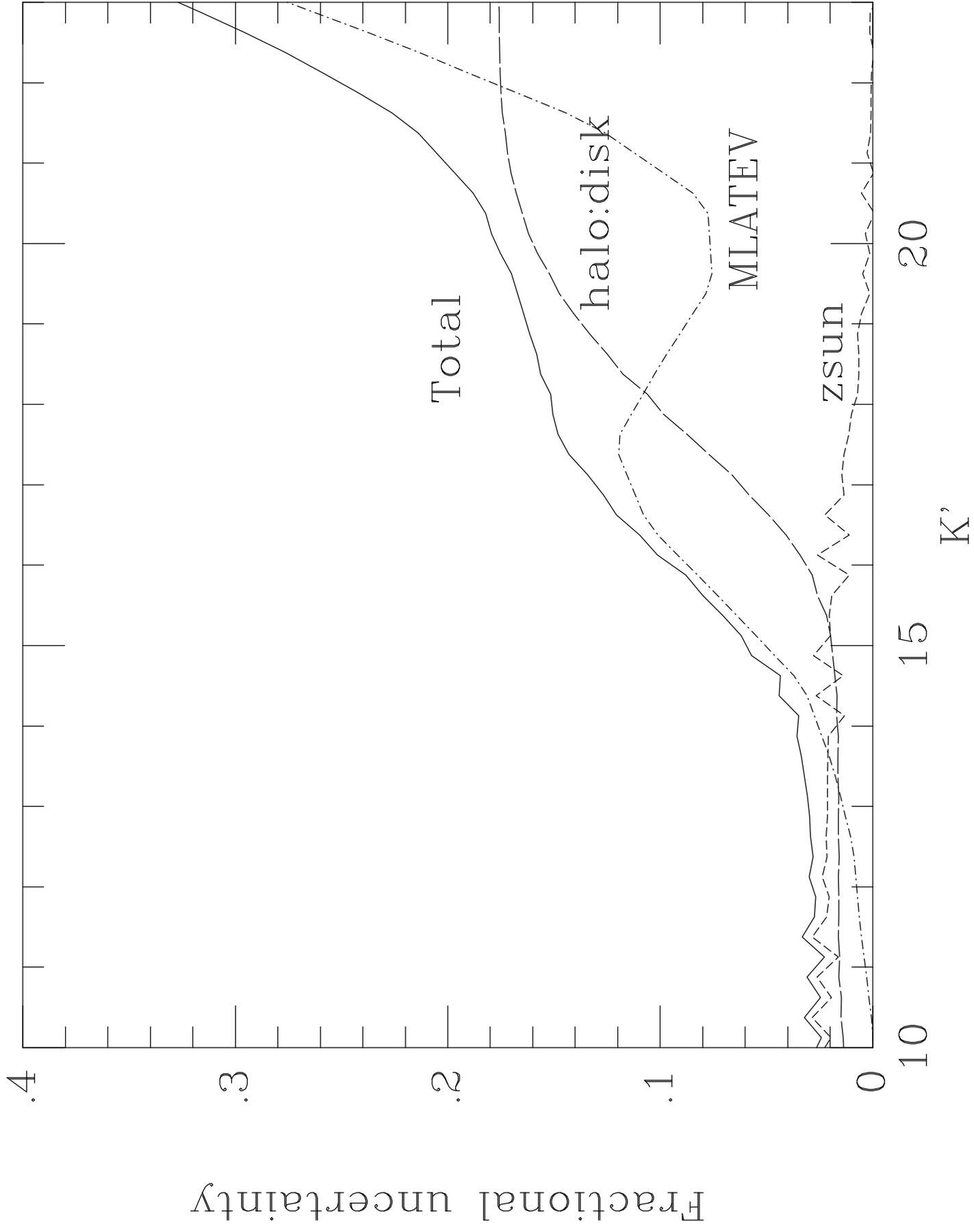
\end{document}